\documentclass[12pt]{article}

\usepackage{amssymb}
\usepackage{dsfont}

\begin{document}

\begin{titlepage}

\begin{flushright}
arXiv:1202.5993
\end{flushright}
\vskip 2.5cm

\begin{center}
{\Large \bf Lorentz Violation and the Higgs Mechanism}
\end{center}

\vspace{1ex}

\begin{center}
{\large Brett Altschul\footnote{{\tt baltschu@physics.sc.edu}}}

\vspace{5mm}
{\sl Department of Physics and Astronomy} \\
{\sl University of South Carolina} \\
{\sl Columbia, SC 29208} \\
\end{center}

\vspace{2.5ex}

\medskip

\centerline {\bf Abstract}

\bigskip

We consider scalar quantum electrodynamics in the Higgs phase
and in the presence of Lorentz violation.
Spontaneous breaking of the gauge symmetry gives rise to Lorentz-violating
gauge field mass
terms. These may cause the longitudinal mode of the gauge
field to propagate superluminally.
The theory may be quantized by the Faddeev-Popov procedure, although the Lagrangian
for the ghost fields also needs to be Lorentz violating.

\bigskip

\end{titlepage}

\newpage

\section{Introduction}

Lorentz violation is currently a topic of significant interest in particle physics
and other areas. No particularly strong evidence for a deviation from
Lorentz invariance has been found, but experimental Lorentz tests are constantly
being refined.
The study of Lorentz symmetry remains an active area of research,
because if any violation of Lorentz invariance
were to be found, that would be a discovery of premier importance.

Violations of Lorentz symmetry may be described in an effective
quantum field theory called
the standard model extension (SME). The SME contains translation-invariant but
Lorentz-violating
corrections to the standard model. These are parameterized by small tensor-valued
background fields~\cite{ref-kost1,ref-kost2}. The most frequently considered
subset of the SME is the
minimal SME, which contains only gauge-invariant,
superficially renormalizable forms of Lorentz
violation. The minimal SME has become the standard framework used for parameterizing
the results of experimental Lorentz tests.

Recent searches for Lorentz violation have included studies of matter-antimatter
asymmetries for trapped charged
particles~\cite{ref-bluhm1,ref-gabirelse,ref-dehmelt1} and bound state
systems~\cite{ref-bluhm3,ref-phillips},
measurements of muon properties~\cite{ref-kost8,ref-hughes}, analyses of
the behavior of spin-polarized matter~\cite{ref-heckel3},
frequency standard comparisons~\cite{ref-berglund,ref-kost6,ref-bear,ref-wolf},
Michelson-Morley experiments with cryogenic
resonators~\cite{ref-muller3,ref-herrmann2,ref-herrmann3,ref-eisele,ref-muller2},
Doppler effect measurements~\cite{ref-saathoff,ref-lane1},
measurements of neutral meson
oscillations~\cite{ref-kost10,ref-kost7,ref-hsiung,ref-abe,
ref-link,ref-aubert}, polarization measurements on the light from cosmological
sources~\cite{ref-carroll2,ref-kost11,ref-kost21,ref-kost22},
high-energy astrophysical
tests~\cite{ref-stecker,ref-jacobson1,ref-altschul6,ref-altschul7,ref-klinkhamer2},
precision tests of gravity~\cite{ref-battat,ref-muller4}, and others.
The results of these experiments set constraints on the various SME coefficients, and
up-to-date information about
most of these constraints may be found in~\cite{ref-tables}.

The one-loop renormalization of various sectors of the minimal SME has
been studied. This has included analyses of
Abelian~\cite{ref-kost3}, non-Abelian~\cite{ref-collad-3}, and
chiral~\cite{ref-collad-1} gauge theories with spinor matter,
as well as scalar field theories
with Yukawa interactions~\cite{ref-ferrero3}. Notably absent from this list
is a full treatment of gauge theories with charged scalar fields. Such theories play
a crucially important role in the standard model, but they are complicated by the
possibility of spontaneous gauge symmetry breaking.

This paper represents a first step towards the understand of Lorentz-violating
scalar quantum electrodynamics (SQED). The emphasis will be on the
Higgs mechanism and the way that it affects the quantization of the theory.
The Higgs mechanism is the most important mechanism for endowing gauge
bosons with mass, because it has a
straightforward generalization to non-Abelian gauge theories.

In Lorentz-invariant SQED, the mass term produced by the Higgs mechanism resembles a Proca mass term. However, if the dynamics of the scalar
field responsible for the gauge symmetry breaking
are not Lorentz invariant, it is possible to have mass terms with
different structures. Any Lorentz violation in the scalar sector will be transferred
to the gauge sector when the Goldstone boson of the spontaneously broken symmetry
is ``eaten'' by the gauge field---becoming the longitudinal component of the
massive vector excitation.
There have been some previous discussions of spontaneous symmetry breaking in the
context of the full electroweak sector of the SME~\cite{ref-kost2,ref-anderson}.
However, earlier work has not focused on how the Lorentz violation affects the
gauge boson mass terms or the quantization of the theory. These will be our primary
objects of study.
Since Lorentz violation is physically a small effect, we shall generally only work to
first order in the SME coefficients.

This paper is organized as follows. In section~\ref{sec-L}, we shall introduce the
SQED Lagrange density with dimensionless Lorentz-violating coefficients. After
including the effects of gauge symmetry breaking, we examine several sectors of the
theory, paying particular attention to the structure of the gauge boson mass terms.
Section~\ref{sec-quant} discusses the quantization of the spontaneously broken gauge
theory, including the introduction of interacting Faddeev-Popov ghosts.
Section~\ref{sec-redef} recasts these results using a change of coordinates, which
can
be used to move certain types of Lorentz violation from one sector of the theory to
another. Finally, section~\ref{sec-concl} summarizes the paper's conclusions.

\section{Lorentz-Violating Lagrangians}
\label{sec-L}

\subsection{Lagrangian Structure}

The Lagrange density for our study of Lorentz-violating SQED is
\begin{equation}
\label{eq-L}
{\cal L}=-\frac{1}{4}F^{\mu\nu}F_{\mu\nu}-\frac{1}{4}k_{F}^{\mu\nu\rho\sigma}
F_{\mu\nu}F_{\rho\sigma}+\left(g^{\mu\nu}+k_{\Phi}^{\mu\nu}\right)
\left(D_{\mu}\Phi\right)^{*}\left(D_{\nu}\Phi\right)+\mu^{2}\Phi^{*}\Phi-
\frac{\lambda}{2}\left(\Phi^{*}\Phi\right)^{2}.
\end{equation}
$D_{\mu}=\partial_{\mu}+ieA_{\mu}$ is the usual covariant derivative, and
$V(\Phi)=-\mu^{2}\Phi^{*}\Phi+\frac{\lambda}{2}\left(\Phi^{*}\Phi\right)^{2}$ is
the scalar potential. The Lorentz violation enters through the
coefficients $k_{F}$ in the gauge sector and $k_{\Phi}$ in the scalar sector. Both
of these background tensors are dimensionless.

There are potentially also CPT-odd operators in both the scalar and vector sectors.
However, the CPT-odd
scalar coefficients $a_{\Phi}^{\mu}$ are unobservable in a theory
with only a
single species of charged matter; they can be eliminated by a redefinition of the
phase of the matter field. The gauge coefficients $k_{AF}^{\mu}$ are not so trivial;
they generate
birefringence in the gauge field propagation and may actually
destabilize the theory. However, this birefringence effect is extremely tightly
constrained; moreover, the $k_{AF}$ does not
interact with the Higgs mechanism in any particularly interesting fashion, and so
$k_{AF}$ will be neglected.

Any other Lorentz-violating terms
constructed from the gauge and scalar fields would need to possess at least one of
the following
undesirable features~\cite{ref-kost2}:  explicit spacetime dependence (with an
accompanying violation of energy-momentum conservation), non-renormalizability
(leading to suppression at low energies), gauge non-invariance (and non-conservation
of the gauge charges), or non-locality (threatening the unitarity of the theory). We
shall not consider such terms, and this ensures that the scalar self-interaction term
must take its usual, Lorentz-invariant form.

$k_{F}$ has the symmetries of the
Riemann tensor and a vanishing double trace, but the structure of of $k_{\Phi}$ is
more subtle.
Reality of the action requires that $k^{\mu\nu}_{\Phi}=k_{S}^{\mu\nu}+
ik_{A}^{\mu\nu}$, where $k_{S}^{\mu\nu}=k_{S}^{\nu\mu}$ is symmetric and traceless
in its
Lorentz indices, while $k_{A}^{\mu\nu}=-k_{A}^{\nu\mu}$ is antisymmetric. The
discrete symmetries of $k_{S}$ are quite similar to those of $k_{F}$.
In a Lorentz-violating
theory, the three spatial reflections that together constitute P are generally
inequivalent. Components of the tensors
$k_{F}$ and $k_{S}$ are odd under a reversal of a specific
spacetime coordinate if that coordinate appear as an Lorentz index an odd number of
times. Overall, a particular coefficient $k_{F}^{\mu\nu\rho\sigma}$ acquires a
sign $(-1)^{\mu}(-1)^{\nu}(-1)^{\rho}(-1)^{\sigma}$ under either a P or T
transformation, where $(-1)^{\mu}=1$ if $\mu=0$ and $(-1)^{\mu}=-1$ if $\mu=1$, 2,
or 3. The transformation of $k_{S}^{\mu\nu}$ is similarly associated with the
sign $(-1)^{\mu}(-1)^{\nu}$.
Both $k_{F}$ and $k_{S}$ are even under charge conjugation (C) and the
combined operation CPT.

However, $k_{A}$ has a different symmetry structure.
$k_{A}^{\mu\nu}$ transforms as $(-1)^{\mu}(-1)^{\nu}$ under P, but the
additional factor of $i$ changes the transformation under C and T.
Under T it transforms as $-(-1)^{\mu}(-1)^{\nu}$,
because T is anti-linear;
$k_{A}$ is also odd under C.
Through integration by
parts, the $k_{A}$ term in ${\cal L}$ is equivalent to
$\frac{1}{2}ek_{A}^{\mu\nu}\Phi^{*}\Phi F_{\mu\nu}$; since $\Phi^{*}\Phi$ is even under C, P, and T separately, this shows that $k_{A}^{\mu\nu}$
must have the same symmetries as $F_{\mu\nu}$.
The fact that the $k_{A}$ term in a minimally coupled but Lorentz-violating
${\cal L}$ can be written in this form means that any additional, non-minimal,
dimension-4 couplings between the scalar and gauge fields are redundant;
their effects are
already completely contained in $k_{A}$. This redundancy was not recognized
in~\cite{ref-anderson}, which studied
$ik_{1}^{\mu\nu}[(D_{\mu}\Phi)^{*}(D_{\nu}\Phi)-(D_{\mu}\Phi)(D_{\nu}\Phi)^{*}]$
and $k_{2}^{\mu\nu}\Phi^{*}\Phi F_{\mu\nu}$ forms of Lorentz violation
separately.

\subsection{Spontaneous Lorentz Violation}

The focus of this paper is primarily on spontaneous {\em gauge} symmetry breaking.
However, the Lorentz-violating terms $k_{F}$, $k_{S}$, and $k_{A}$ could also arise
from spontaneous breaking of Lorentz symmetry. Spontaneous Lorentz breaking has many
advantageous features as a way of introducing Lorentz violating into a theory.
In particular, theories with spontaneous Lorentz violation are consistent with a
pseudo-Riemannian geometric interpretation of gravitation, while explicitly
Lorentz-violating theories generally are not.

The different Lorentz-violating coefficients might be generated by different forms of
Lorentz symmetry breaking. If a symmetric two-index tensor field $X^{\mu\nu}$ has
Lagrange density
\begin{equation}
{\cal L}_{X}=K_{X}(\partial^{\alpha}X^{\mu\nu})-V_{X}(X^{\mu\nu}X_{\mu\nu}),
\end{equation}
with kinetic term $K_{X}$ and potential $V_{X}$, Lorentz symmetry will be broken
if $V_{X}(\zeta)$ has a minimum for either $\zeta>0$ or
$\zeta<0$~\cite{ref-bluhm5,ref-kost25}. The
Lorentz-violating vacuum expectation value of $X_{\mu\nu}$ is $x^{\mu\nu}$, and if
$X^{\mu\nu}$ is coupled to other fields, $x^{\mu\nu}$ can generate either a $k_{S}$ or
$k_{F}$ term. An interaction Lagrange density
$g_{S}X^{\mu\nu}\left(D_{\mu}\Phi\right)^{*}\left(D_{\nu}\Phi\right)$ produces a
$k_{S}$, and $g_{F}X^{\mu\nu}F_{\alpha\mu}F^{\alpha}\,_{\nu}$ a $k_{F}$.
If there is only a single Lorentz-violating vacuum expectation value $x^{\mu\nu}$,
the $k_{S}$ and $k_{F}$ terms may still have different magnitudes,
if the couplings $g_{S}$ and $g_{F}$ differ.

The novel interactions above
also include couplings of $\Phi$ and $A$ to the part of $X^{\mu\nu}$ that
represents fluctuations around the vacuum value $x^{\mu\nu}$. However, these
interactions do not affect any of this paper's results concerning the propagation
modes of the gauge and Higgs fields. Moreover, the interaction vertices involved are
higher dimensional and thus generically suppressed.

A $k_{A}$ term may also be generated if an antisymmetric two-index field
$Y^{\mu\nu}$ with Lagrange density
\begin{equation}
{\cal L}_{Y}=k_{Y}(\partial^{\alpha}Y^{\mu\nu})-V_{Y}(Y^{\mu\nu}Y_{\mu\nu})
\end{equation}
gets a vacuum expectation $y^{\mu\nu}$~\cite{ref-altschul27}.
A coupling
$ig_{A}Y^{\mu\nu}\left[\left(D_{\mu}\Phi\right)^{*}\left(D_{\nu}\Phi\right)-
\left(D_{\mu}\Phi\right)\left(D_{\nu}\Phi\right)^{*}\right]$ then
produces the $k_{A}$
term. (In fact, the antisymmetric $y^{\mu\nu}$ could also generate a
$k_{S}^{\mu\nu}$ proportional to $y^{\mu\alpha}y^{\nu}\,_{\alpha}$, if the theory
includes higher-order couplings.)

\subsection{Spontaneous Gauge Symmetry Breaking}

Whether the Lorentz violation derives from spontaneous Lorentz symmetry breaking or
some other mechanism is an important question. However, the answer has little
direct bearing on the structure of the Higgs sector of Lorentz-violating SQED. We
shall now turn our attention to the systematics of the Lorentz-violating Higgs
mechanism.

Because of the ``wrong-sign'' scalar mass term in (\ref{eq-L}), there are static
solutions to the field equations with nonzero values of $\Phi$.
The Lorentz violation, which appears only in the kinetic terms,
does not affect these solutions, which are derived from
\begin{equation}
\left.\frac{\delta{\cal L}}{\delta\Phi^{*}}\right|_{static}=\mu^{2}\Phi-
\lambda\left(\Phi^{*}\Phi\right)\Phi=0.
\end{equation}
The static solutions $\Phi_{0}$ must satisfy $|\Phi_{0}|=\frac{\mu}{\sqrt{\lambda}}
\equiv v$. Such solutions obviously break the $U(1)$ gauge invariance associated
with the gauge transformation
\begin{eqnarray}
\label{eq-Phi1trans}
\Phi & \rightarrow  & \Phi'=e^{i\alpha}\Phi \\
\label{eq-Phi2trans}
\Phi^{*} & \rightarrow &\Phi'^{*}=e^{-i\alpha}\Phi^{*} \\
\label{eq-Atrans}
A_{\mu} & \rightarrow & A_{\mu}'=A_{\mu}-\frac{1}{e}\partial_{\mu}\alpha.
\end{eqnarray} 
However, it is possible to make $\Phi_{0}$ real by a gauge rotation and then
decompose the field into its vacuum expectation value and excitations,
\begin{equation}
\Phi=v+\frac{1}{\sqrt{2}}(h+i\varphi);
\end{equation}
$h$ is the Higgs field and $\varphi$ represents the Goldstone boson.

The original Lagrange density
${\cal L}$ may be expanded in terms of these new variables, giving
\begin{eqnarray}
{\cal L} & = & -\frac{1}{4}F^{\mu\nu}F_{\mu\nu}-\frac{1}{4}k_{F}^{\mu\nu\rho\sigma}
F_{\mu\nu}F_{\rho\sigma}+\frac{1}{2}\left(g^{\mu\nu}+k_{\Phi}^{\mu\nu}\right)
\Bigl\{(\partial_{\mu}h)(\partial_{\nu}h)+(\partial_{\mu}\varphi)(\partial_{\nu}
\varphi)+e^{2}h^{2}A_{\mu}A_{\nu} \nonumber\\
& & +e^{2}\varphi^{2}A_{\mu}A_{\nu}
+2\sqrt{2}e^{2}vhA_{\mu}A_{\nu}+2e^{2}v^{2}A_{\mu}A_{\nu}
+i[(\partial_{\mu}h)(\partial_{\nu}\varphi)-(\partial_{\mu}\varphi)
(\partial_{\nu}h)] \nonumber\\
& & +i\sqrt{2}ev[(\partial_{\mu}h)A_{\nu}-A_{\mu}(\partial_{\nu}h)]
 +ie[(\partial_{\mu}h)(A_{\nu}h)-(A_{\mu}h)(\partial_{\nu}h)] \nonumber\\
& &
+ie[(\partial_{\mu}\varphi)(A_{\nu}\varphi)-(A_{\mu}\varphi)(\partial_{\nu}\varphi)]
-e[(\partial_{\mu}h)(A_{\nu}\varphi)+(A_{\mu}\varphi)(\partial_{\nu}h)] \nonumber\\
& & +\sqrt{2}ev[A_{\mu}(\partial_{\nu}\varphi)+(\partial_{\mu}\varphi)A_{\nu}]
+e[(A_{\mu}h)(\partial_{\nu}\varphi)+(\partial_{\mu}\varphi)(A_{\nu}h)]\Bigr\}
-V(h,\varphi).
\end{eqnarray}
The expansion of the potential around $v$ takes the standard form,
\begin{equation}
V(h,\varphi)=-\frac{\mu^{4}}{\lambda}+\mu^{2}h^{2}+\mu\sqrt{\frac{\lambda}{2}}h
(h^{2}+\varphi^{2})+\frac{\lambda}{8}(h^{2}+\varphi^{2})^{2}.
\end{equation}

The physical excitations of the theory are the (massive) gauge field
$A$ and the Higgs $h$. Note that because $h$ and $A$ are, respectively, even and odd
under C, mixing between propagation states of these fields must involve the C-odd
coefficient $k_{A}$.

The Goldstone boson field $\varphi$ does not have physical excitations. By working
in the unitarity gauge, we may choose the gauge parameter $\alpha$ so as to make
$\Phi$ everywhere real (at the classical level). This eliminates $\varphi$ from
external states. However, quantum fluctuations in the $\varphi$ field cannot be
entirely eliminated, and the Goldstone boson field will appear as a virtual
intermediary in loop calculations; this is actually crucial to the renormalizability
and unitarity of SQED.

\subsection{Propagation and Interactions}
\label{sec-prop}

Propagation of physical fields
is governed by the portion ${\cal L}_{2}'={\cal L}_{2,Ah}$ of ${\cal L}$ that is
bilinear in just $A$ and $h$. This is
\begin{eqnarray}
\label{eq-L2}
{\cal L}_{2}' & = &
-\frac{1}{4}F^{\mu\nu}F_{\mu\nu}-\frac{1}{4}k_{F}^{\mu\nu\rho\sigma}
F_{\mu\nu}F_{\rho\sigma}+\frac{1}{2}\left(g^{\mu\nu}+k_{S}^{\mu\nu}\right)
(\sqrt{2}ev)^{2}A_{\mu}A_{\nu} \nonumber\\
& & +\frac{1}{2}\left(g^{\mu\nu}+k_{S}^{\mu\nu}\right)
(\partial_{\mu}h)(\partial_{\nu}h)-\frac{1}{2}(\sqrt{2}\mu)^{2}h^{2}
+\frac{1}{\sqrt{2}}evk_{A}^{\mu\nu}hF_{\mu\nu}.
\end{eqnarray}
To the extent that the longitudinal component of the massive gauge field $A$ is
really
the Goldstone boson of the broken symmetry, we should expect
that the longitudinal $A$ should propagate like $\Phi$.
As we shall see at the end of this section, the longitudinal part of $A$ does indeed
propagate like $\Phi$ at high energies, although
this phenomenon is not evident from a
naive inspection of the Lagrange density (\ref{eq-L2}).

However, before addressing this point, we shall show how the mass and kinetic parts
of the full bilinear Lagrange density ${\cal L}_{2}$,
which includes the Goldstone bosons, combine to preserve the
transversality of the gauge propagator. The gauge part of ${\cal L}_{2}$
\begin{equation}
{\cal L}_{2,A}=-\frac{1}{4}F^{\mu\nu}F_{\mu\nu}-\frac{1}{4}k_{F}^{\mu\nu\rho\sigma}
F_{\mu\nu}F_{\rho\sigma}+\frac{1}{2}\left(g^{\mu\nu}+k_{S}^{\mu\nu}\right)
\left(\sqrt{2}ev\right)^{2}A_{\mu}A_{\nu}
\end{equation}
is manifestly transverse, except for the term with the gauge boson mass
$m_{A}=\sqrt{2}ev$, which is certainly not. However, there is also a vertex that
mixes the gauge and Goldstone boson propagators; it comes from
\begin{equation}
{\cal L}_{2,A\varphi}={\cal L}_{2,A}+\frac{1}{2}\left(g^{\mu\nu}+k_{S}^{\mu\nu}
\right)\left\{(\partial_{\mu}\varphi)(\partial_{\nu}\varphi)+m_{A}
[A_{\mu}(\partial_{\nu}\varphi)+(\partial_{\mu}\varphi)A_{\nu}]\right\}.
\end{equation}
To second order in $m_{A}$ and first order in $k_{S}$, there are two possible
insertions from ${\cal L}_{2,A\varphi}$ that contribute to the polarization tensor
$i\Pi^{\mu\nu}(q)$. The first is the photon mass insertion, which contributes
$im_{A}^{2}(g^{\mu\nu}+k_{S}^{\mu\nu})$. The second insertion involves two
$A$-$\varphi$ vertices, with a $\varphi$ propagator between them. This
propagator is
\begin{equation}
D_{\varphi}(q)=\frac{i}{q^{2}}\left(1-k_{S}^{\gamma\delta}\frac{q_{\gamma}
q_{\delta}}{q^{2}}\right),
\end{equation}
making the polarization tensor
\begin{eqnarray}
i\Pi^{\mu\nu}(q) & = & im_{A}^{2}(g^{\mu\nu}+k_{S}^{\mu\nu}) \nonumber\\
& & +\left[m_{A}
(g^{\mu\alpha}+k_{S}^{\mu\alpha})q_{\alpha}\right]\left[
\frac{i}{q^{2}}\left(1-k_{S}^{\gamma\delta}\frac{q_{\gamma}q_{\delta}}
{q^{2}}\right)\right]\left[m_{A}
(g^{\beta\nu}+k_{S}^{\beta\nu})(-q_{\beta})\right] \\
\label{eq-Pi}
& = & im_{A}^{2}\left[g^{\mu\nu}-\frac{q^{\mu}q^{\nu}}{q^{2}}+k_{S}^{\mu\nu}-
k_{S}^{\mu\alpha}\frac{q_{\alpha}q^{\nu}}{q^{2}}-k_{S}^{\beta\nu}\frac{q^{\mu}
q^{\beta}}{q^{2}}+k_{S}^{\gamma\delta}\frac{q^{\mu}q^{\nu}q_{\gamma}q_{\delta}}
{(q^{2})^{2}}\right].
\end{eqnarray}
Although its structure is rather complicated, this tensor is transverse,
$q_{\mu}\Pi^{\mu\nu}=0$. This is a key consistency condition for the theory; it
ensures the conservation of the total charge (including the charge present in the
vacuum).

There are also terms in ${\cal L}_{2}$ that mix $h$ with the other fields.
However, they are less important, for two separate reasons. An insertion with an
intermediate Higgs involves a massive propagator; without a pole at $q^{2}=0$,
this cannot affect the pole structure of the gauge propagator. Moreover, any mixing
of $h$ with $A$ or $\varphi$ violates C. Since $k_{A}$ is the only source of C
violation in the theory, any modification of the $A$ or $\varphi$ propagator by a
virtual $h$ insertion will necessarily be second order in the Lorentz violation.

Special examples of
Lorentz-violating mass terms of the general form
\begin{equation}M^{\mu\nu}A_{\mu}A_{\nu}=\frac{1}{2}(g^{\mu\nu}+
k_{S}^{\mu\nu})m_{A}^{2}A_{\mu}A_{\nu}
\end{equation}
have previously been studied. These mass terms were not considered in the context of
the Higgs mechanism, but the earlier studies' conclusions about photon propagation
remain valid even in the Higgs phase. Mass terms considered
have included an isotropic but boost-in\-var\-i\-ance-violating
$-\frac{1}{2}m_{\gamma}^{2}A_{j}A_{j}$, as an alternative to the Proca mass
term~\cite{ref-gabadadze,ref-dvali}; or
$-\frac{e^{2}}{24\pi^{2}}(b^{2}g^{\mu\nu}+2b^{\mu}b^{\nu})$,
which could be generated by unusual radiative corrections~\cite{ref-altschul8}.
(Note however, that while the
gauge boson mass in these situations is assumed to be small,
the Lorentz-violating and Lorentz-invariant parts of the mass term are of 
comparable size.) Most recently, Lorentz-violating Stueckelberg mass terms
have also been considered~\cite{ref-cambiaso}.
The previous analyses of these models have demonstrated an interesting interplay between Lorentz-violating mass terms and
the the kinetic part of the gauge-sector Lagrangian. In the concrete
examples that were considered in~\cite{ref-gabadadze,ref-dvali,ref-altschul8}, there
were only two distinct eigenvalues in the mass squared matrix
$M^{\mu}\,_{\nu}$. If the eigenvalue $\frac{1}{2}m_{0}^{2}$
corresponding to the timelike direction is smaller in magnitude than a spacelike
eigenvalue $\frac{1}{2}m_{1}^{2}$, there could be propagation with signal and group
velocities greater than 1 and as large as $\frac{m_{1}}{m_{0}}$. However, this
superluminal propagation is limited to modes that are approximately longitudinal.

This shows that
the existence of a Lorentz-violating mass term can have profound effects
on the propagation of gauge bosons, even when the bosons' momenta are far above the
mass scale $m_{A}$.
The mass term (which might be expected to be important
only in the infrared) affects the ultraviolet behavior of the theory through its
influence on the gauge. Requiring charge conservation forces $A$ to obey a gauge
condition $M^{\mu\nu}\partial_{\mu}A_{\nu}=0$. The relative sizes of
the elements of the mass matrix $M^{\mu}\,_{\nu}$ determine the required gauge.
However, the absolute magnitude of the matrix components are irrelevant;
the gauge condition produced by a mass matrix $\zeta M^{\mu}\,_{\nu}$ is
independent of $\zeta$.
 
When a Lorentz-violating gauge field mass term arises through the Higgs mechanism,
there is a clear physical
mechanism underlying superluminal propagation. If the timelike eigenvalue of
$k_{S}^{\mu}\,_{\nu}$ is $\lambda_{0}$ and the largest spacelike eigenvalue is
$\lambda_{1}>\lambda_{0}$,
the free $\Phi$ field has a kinetic term that supports propagation up to speeds of
$\sqrt{\frac{\lambda_{1}}{\lambda_{0}}}$.
When the Goldstone boson is eaten by the gauge field,
this possibility for superluminal propagation is transferred to the gauge field,
although the Lorentz-violating term that makes this possible is part of the mass term
in ${\cal L}_{A}$, rather than the kinetic term.

In addition to the propagation governed by ${\cal L}_{2}$,
there are also
interaction vertices in the theory. For tree-level calculations, only those vertices
involving $A$ and $h$ are needed. These vertices are given by the interaction
Lagrange density
\begin{equation}
{\cal L}_{I}'=\frac{1}{2}\left(g^{\mu\nu}+k_{S}^{\mu\nu}\right)e^{2}(h^{2}+2vh)
A_{\mu}A_{\nu}
-\frac{1}{2}ek_{A}^{\mu\nu}[h(\partial_{\mu}h)A_{\nu}-h(\partial_{\nu}h)A_{\mu}]
-\mu\sqrt{\frac{\lambda}{2}}h^{3}-\frac{\lambda}{8}h^{4}.
\end{equation}
This includes the usual Higgs self-interaction terms from $\Phi^{4}$ theory, as
well as a seagull vertex (involving two Higgs and two gauge fields) and a related
three-particle vertex with one of the Higgs fields replaced by the vacuum expectation
value $v$. The seagull and three-field vertices have their Lorentz structures
modified by $k_{S}$ in precisely the same way as the gauge boson mass term.

The remaining terms in ${\cal L}_{2}'$ and ${\cal L}_{I}'$ are the
C-violating terms involving $k_{A}$. These
can be expressed in terms of the gauge field strength. The
three-field interaction is
equivalent to $\frac{1}{4}ek_{A}^{\mu\nu}h^{2}F_{\mu\nu}$, and there is
no Lorentz-invariant analogue for such a term. Terms involving $k_{S}$ are similar to
Lorentz-invariant terms, in that they involve replacing the Minkowski metric tensor
$g^{\mu\nu}$ with an arbitrary symmetric $k_{S}^{\mu\nu}$. In contrast, there is no
Lorentz-invariant, antisymmetric, two-index tensor to be contracted with
$F^{\mu\nu}$,
so the $k_{A}^{\mu\nu}F_{\mu\nu}$ interactions have a uniquely Lorentz-violating
structure.

\section{Quantization and Ghost Fields}
\label{sec-quant}

Calculation of quantum corrections for a theory with spontaneously broken gauge
symmetry requires the introduction of a gauge fixing term in the action (which
leads naturally to the inclusion of Faddeev-Popov ghosts). The gauge fixing term
serves two purposes. It can eliminate the zero modes in the gauge field action, which
is necessary for the derivation of a well-defined propagator; the gauge fixing term
fulfills this function in all gauge theories, whether or not they involve
spontaneous symmetry breaking. However, when the gauge symmetry is broken,
the gauge fixing function may also be chosen to remove
any terms that mix the gauge and Goldstone boson fields.

To quantize the gauge field according to the Faddeev-Popov
procedure~\cite{ref-faddeev}, we begin with
the gauge-invariant functional integral for the theory and insert the identity,
in the form
\begin{equation}
\label{eq-ident}
\mathds{1}=\int{\cal D}\alpha(x)\,\delta[G(A',h',\varphi')-\omega]
\det\left[\frac{\delta G(A',h',\varphi')}{\delta\alpha}\right],
\end{equation}
where $A'_{\mu}=A_{\mu}-\frac{1}{e}\partial_{\mu}\alpha$, $h'=h-\alpha\varphi$,
and $\varphi'=\varphi+\alpha(\sqrt{2}v+h)$ are the infinitesimally gauge transformed
fields from (\ref{eq-Phi1trans})--(\ref{eq-Atrans}). The gauge-fixing function
$[G(A,h,\varphi)-\omega]$ is then integrated over a Gaussian distribution of
$\omega$ values.
We take
\begin{equation}
\label{eq-G}
G=\frac{1}{\sqrt{\xi}}\left[\left(g^{\mu\nu}+k_{G}^{\mu\nu}\right)\partial_{\mu}
A_{\nu}-\sqrt{2}\xi ev\varphi\right].
\end{equation}
The Lorentz-invariant terms in (\ref{eq-G}) are identical to those in
the gauge fixing function for the
$R_{\xi}$ gauge,
and $k_{G}$ is an (as yet undetermined) Lorentz-violating tensor coefficient. It is
not possible to include Lorentz violation in the $\varphi$ part of $G$ without
introducing higher derivatives into the final ghost action.

The Faddeev-Popov
procedure introduces two new sets of terms into the Lagrange density. The first
set is the result of the integration over $\omega$,
\begin{equation}
-\frac{1}{2}G^{2}=-\frac{1}{2\xi}
\left[\left(g^{\mu\nu}+k_{G}^{\mu\nu}\right)\partial_{\mu}A_{\nu}\right]^{2}
-\sqrt{2}ev\left(g^{\mu\nu}+k_{G}^{\mu\nu}\right)(\partial_{\mu}\varphi)A_{\nu}-
\xi e^{2}v^{2}\varphi^{2}.
\end{equation}
The Lorentz violation $k_{G}$ in the gauge fixing should be chosen to eliminate
the $A$-$\varphi$ mixing term in ${\cal L}_{2}-\frac{1}{2}G^{2}$. This
requires $k_{G}^{\mu\nu}=k_{S}^{\mu\nu}$, and the gauge part of ${\cal L}_{2}$
becomes

\newpage

\begin{eqnarray}
{\cal L}_{2,A}-\frac{1}{2\xi}
\left[\left(g^{\mu\nu}+k_{G}^{\mu\nu}\right)\partial_{\mu}A_{\nu}\right]^{2} & = &
-\frac{1}{4}F^{\mu\nu}F_{\mu\nu}-\frac{1}{2\xi}(\partial^{\mu}A_{\mu})^{2}
+\frac{1}{2}m_{A}^{2}A^{\mu}A_{\mu} \\
& & -\frac{1}{4}k_{F}^{\mu\nu\rho\sigma}F_{\mu\nu}F_{\rho\sigma}-\frac{1}{\xi}
k_{S}^{\mu\nu}(\partial_{\mu}A_{\nu})(\partial^{\rho}A_{\rho})
+\frac{1}{2}k_{S}^{\mu\nu}m_{A}^{2}A_{\mu}A_{\nu}. \nonumber
\end{eqnarray}
The Lorentz-violating kinetic terms can be
recast as $-k_{\xi}^{\mu\nu\rho\sigma}(\partial_{\mu}A_{\nu})(\partial_{\rho}
A_{\sigma})$, where $k_{\xi}^{\mu\nu\rho\sigma}=k_{F}^{\mu\nu\rho\sigma}+\frac{1}
{\xi}g^{\mu\nu}k_{S}^{\rho\sigma}$.
The $(\partial^{\mu}A_{\mu})^{2}$
term combines with the Maxwell and Proca terms to produce the usual propagator
\begin{equation}
D^{\mu\nu}_{A}(q)=\frac{-i}{q^{2}-m_{A}^{2}}\left[g^{\mu\nu}-(1-\xi)
\frac{q^{\mu}q^{\nu}}{q^{2}-\xi m_{A}^{2}}\right],
\end{equation}
and the Lorentz-violating terms may be treated as vertices. The
$\xi$-dependent part of the $k_{\xi}$ vertex is superficially similar in structure to
the $k_{F}$ part. However, while the $k_{F}$ term
only involves the physical degrees of freedom contained in $F^{\mu\nu}$, the gauge
fixing part only involves purely gauge degrees of freedom, since $k_{G}$ couples to
the symmetric part of $\partial_{\mu}A_{\nu}$.

The other terms that the Faddeev-Popov procedure adds to ${\cal L}$ come from the
functional determinant in (\ref{eq-ident}). Since
\begin{eqnarray}
\frac{\delta G}{\delta\alpha} & = & \frac{\delta G}{\delta A_{\mu}}\left(\frac{1}{e}
\partial_{\mu}\right)+\frac{\delta G}{\delta\varphi}(v+h) \\
& = & \frac{1}{\sqrt{\xi}}\left[\left(g^{\mu\nu}+
k_{S}^{\mu\nu}\right)\left(-\frac{1}{e}\partial_{\mu}\partial_{\nu}\right)-\xi m_{A}
(\sqrt{2}v+h)\right],
\end{eqnarray}
$\det[\delta G/\delta\alpha]$ may be exponentiated as a part of the action by
introducing ghost fields $c$ and $\bar{c}$ with Lagrange density
\begin{equation}
{\cal L}_{c}=\left(g^{\mu\nu}+k_{S}^{\mu\nu}\right)(\partial_{\mu}\bar{c})
(\partial_{\nu}c)-\xi m_{A}^{2}\left(1+\frac{h}{\sqrt{2}v}\right)\bar{c}c.
\end{equation}
The (gauge-dependent) mass term for the Faddeev-Popov ghosts is
unaffected by the Lorentz violation; the interaction vertex with
the Higgs field is also unmodified. However, the ghosts do acquire a modification
to their kinetic term, equivalent to the $k_{S}$ for the original scalar field
$\Phi$. For each of the spinless fields ($h$, $\varphi$, and $c$), the Lorentz
violation may be treated as a vertex to be inserted along propagation lines.
Several loop diagrams involving the $k_{S}$ in the ghost sector have already been
evaluated~\cite{ref-altschul25}.


\section{Coordinate Redefinitions}
\label{sec-redef}

The appearance of the same Lorentz-violating coefficients
$k_{S}$ in the Higgs, Goldstone boson, and ghost sectors may be unsurprising,
because of the structure of the $k_{S}$ term. If $k_{F}$ vanishes in the original
Lagrange density ${\cal L}$, then $k_{S}$ describes a mismatch between the natural
coordinates for describing the gauge and matter fields. Having a vanishing $k_{F}$
means that the chosen coordinates are natural for the gauge field. However,
redefining coordinates according to
\begin{equation}
\label{eq-redef}
x^{\mu}\rightarrow x'^{\mu}=x^{\mu}-\frac{1}{2}k_{S}^{\mu}\,_{\nu}x^{\nu}
\end{equation}
will transform the Lorentz violation coefficients in ${\cal L}$ to
\begin{eqnarray}
k_{\Phi}^{\mu\nu} & \rightarrow & k_{\Phi}'^{\mu\nu}=ik_{A}^{\mu\nu} \\
\label{eq-kF-redef}
k_{F}^{\mu\nu\rho\sigma} & \rightarrow & k_{F}'^{\mu\nu\rho\sigma}=
k_{F}^{\mu\nu\rho\sigma}-
\frac{1}{2}\left(g^{\mu\rho}k_{S}^{\nu\sigma}-g^{\mu\sigma}k_{S}^{\nu\rho}
-g^{\nu\rho}k_{S}^{\mu\sigma}+g^{\nu\sigma}k_{S}^{\mu\rho}\right).
\end{eqnarray}
If this transformation is made prior to the calculations, many of the
Lorentz-violating terms that could appear after spontaneous symmetry breaking are
actually absent. By eliminating $k_{S}$ prior to quantization and spontaneous
symmetry breaking, we can ensure that there is no Lorentz-violating modification of
the gauge field mass term, nor is any Lorentz violation required in the ghost sector.

In fact,
it is straightforward to see how a transformation that eliminates $k_{S}$ from
the kinetic term for $\Phi$ likewise eliminates $k_{S}$ from the ghost kinetic
term. Both terms have the same basic scalar kinetic structure, and a transformation
that carries $(g^{\mu\nu}+k_{S}^{\mu\nu})\partial_{\mu}\partial_{\nu}\rightarrow
\partial^{\mu}\partial_{\mu}$ will have the same effect in either sector.
Accompanying
the redefinition of the coordinates (\ref{eq-redef}) must be a similar linear
reshuffling of the gauge fields; the transformed $A'_{\mu}$ must be exactly what
enters in conjunction with $\partial'_{\mu}\equiv\frac{\partial}{\partial x'^{\mu}}$
in the covariant derivative. So (\ref{eq-redef}) simply takes
$(g^{\mu\nu}+k_{S}^{\mu\nu})A_{\mu}A_{\nu}\rightarrow
A^{\mu}A_{\mu}$.


Since it is possible to define away the $k_{S}$ Lorentz violation, we might be
tempted to dismiss analyses that include $k_{S}$ entirely as
unnecessary.
However, since the transformation
that eliminates $k_{S}$ is a global redefinition of the coordinates, it can only be
used to eliminate this type of Lorentz violation from a single sector. This is
already evident from the fact that removing $k_{S}$ from the matter sector introduces
it into the $k_{F}'$ of the gauge sector. Ultimately, physical observables that
depend on $k_{S}$ need to involve differences between SME coefficients across
different sectors. In pure SQED, the only observable
difference is $k_{S}^{\mu\nu}-k_{F\alpha}\,^{\mu\alpha\nu}$.

If both the Lorentz violation and the mass are small enough to be treated as
perturbations, it is straightforward to determine the dispersion relations for the
gauge field modes in the coordinate system with all Lorentz violation moved into the
gauge sector. For the transverse polarization states with wave vector $\vec{q}$,
the frequencies are~\cite{ref-kost16}
\begin{equation}
\label{eq-omega}
q_{0}^{\pm}=\left|\vec{q}\,\right|\left[1+\rho\left(\hat{q}\right)\pm\sigma
\left(\hat{q}\right)\right]+\frac{m_{A}^{2}}{2\left|\vec{q}\,\right|},
\end{equation}
where $\rho\left(\hat{q}\right)=-\frac{1}{2}\tilde{k}^{\alpha}\,_{\alpha}$, and
$\sigma^{2}\left(\hat{q}\right)=\frac{1}{2}\tilde{k}^{\alpha\beta}\tilde{k}
_{\alpha\beta}-\rho^{2}\left(\hat{q}\right)$, with $\tilde{k}^{\alpha\beta}=
k_{F}'^{\alpha\mu\beta\nu}\hat{q}_{\mu}\hat{q}_{\nu}$ and $\hat{q}^{\mu}=
\left(1,\vec{q}/\left|\vec{q}\,\right|\right)$.
The result (\ref{eq-omega}) simply represents the conventional dispersion
relation, plus the usual perturbations due to the $k_{F}'$ Lorentz violation
and the mass $m_{A}\ll\left|\vec{q}\,\right|$.

However, there is also a
longitudinal polarization state, whose energy is not affected by $k_{F}'$ at leading
order,
\begin{equation}
\label{eq-omega2}
q_{0}=\left|\vec{q}\,\right|+\frac{m_{A}^{2}}{2\left|\vec{q}\,\right|}.
\end{equation}
The reason that $k_{F}'$ does not affect this dispersion relation is that
the presence of the mass term forces $A$ to obey the Lorenz gauge condition
$q^{\mu}A_{\mu}=0$ (plus Lorentz-violating corrections that may be neglected at this
order).
This makes the $k_{F}'$ term in the equation of motion for the longitudinal mode
vanish identically. The lack of any dependence on $k_{F}'$ might initially seem puzzling, but
it is actually quite natural. Since $m_{A}^{2}$ was treated as a perturbation,
(\ref{eq-omega2}) applies only in the high-energy regime, when the momentum
$\left|\vec{q}\,\right|$ is large compared with the Higgs mass scale. In that regime,
the longitudinal component of the gauge field essentially becomes indistinguishable
from the uneaten
Goldstone boson. The propagation of the longitudinal mode should therefore
be governed only by the Lorentz-violating tensor $k_{S}'$ in the Higgs sector, and in
the transformed coordinates used to derive (\ref{eq-omega2}), $k_{S}'$ vanishes.

Of course,
these dispersion relations may be transformed back into the original coordinates
with nonzero $k_{S}$ simply by inverting the coordinate redefinition
(\ref{eq-redef}), so that $q^{\mu}\rightarrow q^{\mu}-
\frac{1}{2}k_{S}^{\mu}\,_{\nu}q^{\nu}$. The result for the longitudinal mode is
\begin{equation}
q_{0}=\left|\vec{q}\,\right|\left[1-\frac{1}{2}k_{S}^{00}-2k_{S}^{0}\,_{j}\hat{q}_{j}
+\frac{1}{2}k_{S}^{j}\,_{l}
\hat{q}_{j}\hat{q}_{l}\right]
+\frac{m_{A}^{2}}{2\left|\vec{q}\,\right|}
\end{equation}
This exhibits exactly the same kind of potentially superluminal behavior for the
longitudinal mode as was discussed in section~\ref{sec-prop}, with the limiting
speed controlled by the relative sizes of the spacelike and timelike eigenvalues of
$k_{S}^{\mu}\,_{\nu}$.

This analysis also provides insight into another feature of the non-Higgs mass models
discussed in~\cite{ref-altschul8}. The normal modes of propagation in the
presence of the Lorentz-violating mass term do not involve orthogonal polarization
vectors. This is related to the non-orthogonal nature of the transformation
(\ref{eq-redef}); a coordinate redefinition that moves Lorentz violation from the
gauge field kinetic term to the mass term changes an orthogonal basis of polarization
states into a non-orthogonal one. In fact, the transformation required to turn a
Lorentz-invariant Proca mass term into the term
$-\frac{e^{2}}{24\pi^{2}}(b^{2}g^{\mu\nu}+2b^{\mu}b^{\nu})$ from~\cite{ref-altschul8}
would produce extremely skewed coordinates. This is a reminder that, while the gauge
boson mass parameters
in~\cite{ref-gabadadze,ref-dvali,ref-altschul8} may be small, the Lorentz violation
for the theories involved is, in a meaningful sense, quite large---with the
equivalent of $k_{S}$ being ${\cal O}(1)$.

\section{Conclusion}
\label{sec-concl}

The focus of this paper has been on SQED with Lorentz violation. With a single
scalar field $\Phi$ and a single gauge field $A$, all possible
forms of renormalizable, CPT-even Lorentz
violation are captured in the coefficients $k_{\Phi}$ and $k_{F}$, with minimal
coupling of the gauge and matter fields through the covariant derivative $D_{\mu}$.
In standard SQED, spontaneous breaking of the $U(1)$ gauge symmetry makes the
gauge boson massive. We have shown that the Lorentz-violating theory includes an
analogous mass term, with a Lorentz-violating generalization of the Proca form.

We have displayed the full Lagrange density for this theory and for the first time
introduced the Faddeev-Popov ghosts that are a necessary part of the quantization
procedure. The effects of Lorentz violation on the ghosts has already been studied
for non-Abelian gauge theories~\cite{ref-collad-3}, but not for theories with a
broken gauge symmetry. Knowledge of the full Faddeev-Popov Lagrange density will make
it possible to perform Feynman diagram calculations in the present theory.

We have also shown how Lorentz violation in the scalar and gauge
sectors affects the propagation of the physical gauge and Higgs modes. Even with a conventional kinetic term $-\frac{1}{4}F^{\mu\nu}F_{\mu\nu}$ for the gauge field,
it may be possible for the longitudinal mode to propagate superluminally. This can
happen because the longitudinal mode is really an ``eaten'' Goldstone boson, whose
behavior is primarily governed by the structure of the scalar sector.

Quantum field theories involving gauge interactions with charged scalar matter
are important; in the standard model, the Higgs sector is responsible for the
existence of particle masses. The full treatment of quantum corrections in the SME
is an interesting theoretical problem, and this work presents an important step
toward a complete understanding of the SME.

\end{document}